\RequirePackage{lineno} 
\documentclass[aps,prc,superscriptaddress,showpacs,floatfix,nofootinbib,onecolumn]{revtex4}
\usepackage{graphicx} 
\usepackage{float} 
\usepackage{amssymb}
\usepackage{url}
\usepackage{harpoon}
\usepackage{amsmath}
\usepackage{MnSymbol}
\usepackage{multirow}
\usepackage[colorlinks,breaklinks]{hyperref}
\hypersetup{linkcolor=blue,citecolor=blue,filecolor=black,urlcolor=blue}

\newcommand{\pA}{p+{\mathrm{A}}}

\begin{document}
\title{Limitation of multi-particle correlations for studying the event-by-event distribution of harmonic flow in heavy-ion collisions}
\newcommand{\sunysb}{Department of Chemistry, Stony Brook University, Stony Brook, NY 11794, USA}
\newcommand{\bnl}{Physics Department, Brookhaven National Laboratory, Upton, NY 11796, USA}
\author{Jiangyong Jia}\email[Correspond to\ ]{jjia@bnl.gov}
\affiliation{\sunysb}\affiliation{\bnl}
\author{Sooraj Radhakrishnan}\email[Correspond to\ ]{sooraj9286@gmail.com}
\affiliation{\sunysb}

\begin{abstract}
The sensitivity of flow harmonics from cumulants on the event-by-event flow distribution $p(v_n)$ is investigated using a simple central moment expansion approach. For narrow distribution whose width is much smaller than the mean $\sigma_n\ll \left\langle v_n\right\rangle$, the difference between the first three higher-order cumulant estimates $v_n\{4\}$, $v_n\{6\}$ and $v_n\{8\}$ are not very sensitive to the shape of $p(v_n)$. For broad distribution $\sigma_n\gtrsim \left\langle v_n\right\rangle$, the higher-order cumulant estimates differ from each other but may change sign and become ill-defined. This sign change arises from the choice of $p(v_n)$, without the need to invoke non-flow effects. Direct extraction of $p(v_n)$ via a data-driven unfolding method used by the ATLAS experiment is a more preferred approach for flow distribution measurement.
\end{abstract}
 \maketitle 
\section{Introduction} \label{sec:1}
Heavy-ion collisions at RHIC and LHC create a new state of nuclear matter that behaves like a perfect fluid of quarks and gluons (quark-gluon plasma or QGP), characterized by a small $\eta/s$ value (ratio of shear viscosity to entropy density) that is close to the conjectured lower bound~\cite{Heinz:2013th,Gale:2013da}. The properties of the QGP, including $\eta/s$, are often inferred from studies of the collective flow phenomena~\cite{Voloshin:2008dg}. Due to its small value of $\eta/s$, the collective expansion of the QGP efficiently transfer the asymmetries of the initial geometry into the azimuthal anisotropy of produced particles in momentum space. Detailed measurements of the collective flow and successful descriptions by hydrodynamic models have placed important constraints on the transport properties and initial conditions of the QGP~\cite{Luzum:2013yya}.

The azimuthal anisotropy of the particle production in an event can be characterized by Fourier expansion of the underlying probability distribution $\emph{P}(\phi)$ in azimuthal angle $\phi$~\cite{Ollitrault:1992bk,Voloshin:1994mz}, 
\begin{eqnarray}
\label{eq:1}
\emph{P}(\phi) = \frac{1}{2\pi} \sum_{n=-\infty}^{\infty} {\bf v}_{n}e^{-in\phi},\;\;\; {\bf v}_n = v_n e^{in\Phi_n}
\end{eqnarray}
where $v_n$ and $\Phi_n$ are the magnitude and phase of the $n$-th order harmonic flow. Due to large event-by-event (EbyE) fluctuations of the collision geometry in the initial state, the $v_n$ varies event to event, and can be described by a probability distribution $p(v_n)$~\cite{Luzum:2013yya,Jia:2014jca}.

Since the number of measured particles $M$ in each event is finite, the flow vectors ${\bf v}_n$ can only be estimated, for example by:
\begin{eqnarray}
\label{eq:2a}
{\bf q}_n = q_n e^{in\Psi_n} \equiv \frac{\sum_i  e^{in\phi_i}}{M}
\end{eqnarray}
where the sum runs over the particles in the event, $\phi_i$ are their azimuthal angles and $\Psi_n$ is the observed event plane. In the presence of non-flow ${\boldsymbol s}_{n}$ which includes statistical fluctuation and various short-range correlations, the magnitude and direction of ${\bf q}_n$ differ from truth flow:
\begin{eqnarray}
\label{eq:2b}
{\bf q}_n = {\bf v}_{n}+{\boldsymbol s}_{n}
\end{eqnarray}
It is important to emphasize that since non-flow sources are {\it uncorrelated} event to event, the probability distributions for ${\bf q}_n$ and ${\bf v}_{n}$ can be related to each other simply by a random smearing function $p({\boldsymbol s}_{n})$ that reflects the EbyE distribution of non-flow.
\begin{eqnarray}
\label{eq:2c}
p({\bf q}_n) = p({\bf v}_{n})\otimes p({\boldsymbol s}_{n})
\end{eqnarray}
$p({\boldsymbol s}_{n})$ can be calculated from a data-driven unfolding method introduced by the ATLAS Collaboration~\cite{Aad:2013xma}. The key step in this method is to determine a response function, $r(v_n^{{\mathrm{obs}}}|v_n)$, which connects the observed signal $v_n^{{\mathrm{obs}}}$ with the true $v_n$ signal. The response function is then used to unfold the $v_n^{{\mathrm{obs}}}$ distributions to obtain the true $v_n$ distributions, using the Bayesian unfolding technique~\cite{Agostini}. Since response function contains statistical fluctuation and various short-range correlations, they are naturally removed from $v_n$ in the unfolding procedure. Alternatively, $p({\boldsymbol s}_{n})$ can also be estimated from a event generator that does not have collective flow~\cite{Jia:2013tja}, such as HIJING~\cite{Gyulassy:1994ew}. Both studies show that $p({\boldsymbol s}_{n})$ is Gaussian for sufficiently large $M$ (or $t$-distribution for moderate $M$). Once $p({\boldsymbol s}_{n})$ is known, $p({\bf v}_n)$ can be obtained from $p({\bf q}_n)$ via a statistical unfolding method~\cite{Agostini}. The resolving power on the shape of $p(v_n)$ of this method is controlled only by the width of $p({\boldsymbol s}_{n})$. The first result of $p(v_n)$ has been obtained in this way for $n=2$, 3 and 4~\cite{Aad:2013xma}.

A more traditional method to study $p(v_n)$ is using the cumulants from multi-particle correlations~\cite{Borghini:2000sa,Borghini:2001vi,Bilandzic:2010jr,Bilandzic:thesis}. A $2k$-particle azimuthal correlator is obtained by averaging over all unique combinations in one event then over all events:
\begin{eqnarray}
\label{eq:3a}
\left\langle\left\langle 2k\right\rangle\right\rangle = \left\langle\left\langle e^{in\sum_{j=1}^{k}(\phi_{2j-1}-\phi_{2j})}\right\rangle\right\rangle \approx \left\langle q_n^{2k}\right\rangle = \left\langle v_n^{2k}\right\rangle +\mathrm{\mbox{non-flow}}.
\end{eqnarray}
where $\left\langle x^{2k}\right\rangle\equiv\int x^{2k} p(x)dx$ is the $2k$-th moment of the probability distribution for $x$. The cumulants are then obtained by proper combination of all correlations involving $\leq2k$ number of particles. The formulae for the first four cumulants are~\cite{Borghini:2001vi}:
\begin{eqnarray}\nonumber
c_n\{2\} &=& \left\langle\left\langle 2\right\rangle\right\rangle\\\nonumber
c_n\{4\} &=& \left\langle\left\langle 4\right\rangle\right\rangle-2\left\langle\left\langle 2\right\rangle\right\rangle^2\\\nonumber
c_n\{6\} &=& \left\langle\left\langle 6\right\rangle\right\rangle-9\left\langle\left\langle 4\right\rangle\right\rangle\left\langle\left\langle 2\right\rangle\right\rangle+12\left\langle\left\langle 2\right\rangle\right\rangle^3\\\label{eq:3b}
c_n\{8\} &=& \left\langle\left\langle 8\right\rangle\right\rangle-16\left\langle\left\langle 6\right\rangle\right\rangle\left\langle\left\langle 2\right\rangle\right\rangle-18\left\langle\left\langle 4\right\rangle\right\rangle^2+144\left\langle\left\langle 4\right\rangle\right\rangle\left\langle\left\langle 2\right\rangle\right\rangle^2-144\left\langle\left\langle 2\right\rangle\right\rangle^4
\end{eqnarray}
which leads to the following expressions for the harmonic flow $v_n$:
\begin{eqnarray}\label{eq:3c}
v_n\{2\} = \sqrt{c_n\{2\}},\;v_n\{4\} =\sqrt[4]{-c_n\{4\}},\;v_n\{6\} = \sqrt[6]{c_n\{6\}/4},\;v_n\{8\} =\sqrt[8]{-c_n\{8\}/33}
\end{eqnarray}
The cumulant framework has been generalized into all particle correlations known as the lee-yang-zero (LYZ) method~\cite{Bhalerao:2003xf}. 

The main advantage of cumulants is that they enable the subtraction of non-flow effects from genuine flow order by order after ensemble average~\cite{Borghini:2001vi,Bilandzic:2010jr}. But this method measures only the ``even'' moments of $p(v_n)$, and there is no known analytical approach yet that can be used to reliably reconstruct the flow distribution from the first several cumulants (unless the shape of the distribution is known). One interesting question is how sensitive are these higher-order cumulants to the underlying flow fluctuation, and how well one can reconstruct the actual $p(v_n)$ from the first few $v_n\{2k\}$. Answer to this question is especially important with the recent development of event-shape selection technique~\cite{Schukraft:2012ah,Huo:2013qma,Jia:2014ysa} or in the study of flow in collisions of deformed nucleus~\cite{Wang:2014qxa,Rybczynski:2012av}, where the $p(v_n)$ can have very different shapes.

\begin{figure}[!h]
\center
\includegraphics[width=0.7\linewidth]{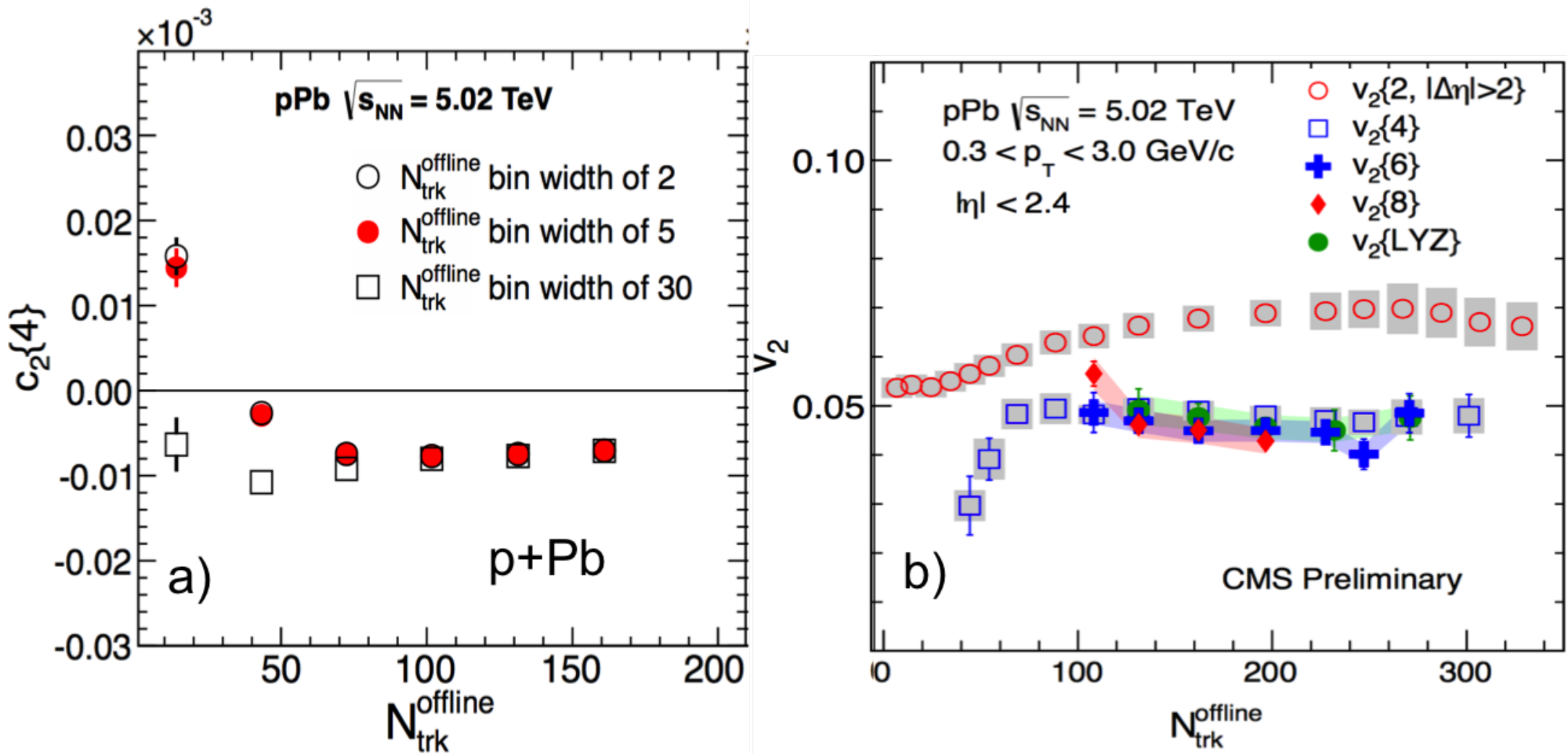}
\caption{\label{fig:0} (Color online) Cumulants for second order azimuthal harmonics as a function of number of charged tracks in $p$+Pb collisions: (a) $c_2\{4\}$ from Ref.~\cite{Chatrchyan:2013nka}, (b) flow harmonics $v_2\{2k\}$ for $k=1$ to 4 from Ref.~\cite{Quan}.}
\end{figure}
There are no simple analytical expression of cumulants for arbitrary probability distribution. However, if the distribution of flow vector is Gaussian or equivalently the distribution of $v_n$ is Bessel-Gaussian
\begin{eqnarray}
\label{eq:4a}
p({\bf v}_n) =\frac{1}{2\pi\delta^2_{n}}e^{-\left|{\bf v}_n-v_n^{\;_0}\right|^2 \big{/}\left(2\delta^2_{n}\right)},\; p(v_n) =\frac{v_n}{\delta_{n}^2}e^{-\frac{(v_n)^2+(v_n^{0})^2}{2\delta_{n}^2}} I_0\left(\frac{v_n^{0}v_n}{\delta_{n}^2}\right),
\end{eqnarray}
flow harmonics defined by cumulants have a simple expression~\cite{Voloshin:2007pc}:
\begin{eqnarray}
\label{eq:4}
v_n\{2k\} &=& \left\{\begin{array}{ll} \sqrt{\left(v_n^{\mathrm{0}}\right)^2+2\delta^2_{n}} & k=1\\
v_n^{\mathrm{0}} & k>1\\   \end{array}\right.
\end{eqnarray}
Based on this, one may conclude that the observation of $v_n\{4\}\approx v_n\{6\}\approx v_n\{8\}$ in A+A collisions~\cite{Aad:2014vba,Bilandzic:2011ww} and high multiplicity $p$+A collisions~\cite{Quan} (see Fig~\ref{fig:0} (b)) is an indication that the underlying $p({\bf v}_n)$ is close to Gaussian. It is also argued that if flow signal dominates, then cumulants $c_n\{2k\}$ must have the ``correct'' sign, i.e. negative for even $k$ and positive for odd $k$, such that $c_2\{4\}$ is positive for small multiplicity events (see Fig.~\ref{fig:0}(a)) must then imply the dominance of non-flow~\cite{Aad:2013fja,Chatrchyan:2013nka,Abelev:2014mda}. In this paper, we investigate the sensitivity of cumulants on $p(v_n)$ using a simple central moment expansion approach. We show both claims are false for arbitrary probability distribution, and in many cases the flow harmonics from higher-order cumulants do not provide strong constrains on the shape of $p(v_n)$ beyond its mean and width. Instead, inferring the shape information of $p(v_n)$ directly from the EbyE distribution of observed flow vector $p({\bf q}_n)$ and non-flow $p({\boldsymbol s}_{n})$ is a better approach.

\section{Behavior of cumulants for narrow distributions}\label{sec:2}
In the limit of large $M$, the cumulants are fully determined from the moments of the underlying flow probability distribution (eqs.~\ref{eq:3a} and \ref{eq:3b}):
\begin{eqnarray}\nonumber
\mathcal{C}_n\{2\}  \equiv c_n\{2\}&=& \left\langle v_n^2\right\rangle\\\nonumber
-\mathcal{C}_n\{4\} \equiv c_n\{4\}&=&\left\langle v_n^4\right\rangle-2\left\langle v_n^2\right\rangle^2\\\nonumber
4\mathcal{C}_n\{6\} \equiv c_n\{6\}&=&\left\langle v_n^6\right\rangle-9\left\langle v_n^4\right\rangle\left\langle v_n^2\right\rangle+12\left\langle v_n^2\right\rangle^3\\\label{eq:3d}
-33\mathcal{C}_n\{8\} \equiv c_n\{8\}&=&\left\langle v_n^8\right\rangle-16\left\langle v_n^6\right\rangle\left\langle v_n^2\right\rangle-18\left\langle v_n^4\right\rangle^2+144\left\langle v_n^4\right\rangle\left\langle v_n^2\right\rangle^2-144\left\langle v_n^2\right\rangle^4.
\end{eqnarray}
where new variables $\mathcal{C}_n\{2k\}$ are introduced in order to have a unified expression for the flow coefficients (and extended to negative values):
\begin{eqnarray}
\label{eq:3e}
v_n\{2k\} &\equiv& \mbox{sgn}(\mathcal{C}_n\{2k\})\sqrt[2k]{|\mathcal{C}_n\{2k\}|}
\end{eqnarray}

The moment of the flow distribution can be expanded into a finite number of central moments:
\begin{eqnarray}
\label{eq:5a}
\left\langle v_n^{2k}\right\rangle = \left\langle v_n\right\rangle^{2k} \int \left(1+\frac{v_n-\left\langle v_n\right\rangle}{\left\langle v_n\right\rangle}\right)^{2k} p(v_n) dv_n= \left\langle v_n\right\rangle^{2k} \left(1+\sum_{j=2}^{2k}C_{2k}^{j} \theta_j\right),
\end{eqnarray}
where $\theta_j= \int \left(\frac{v_n-\left\langle v_n\right\rangle}{\left\langle v_n\right\rangle}\right)^{j} p(v_n) dv_n$ is the central moment normalized by the $j$-th power of the mean $\left\langle v_n\right\rangle$ (or reduced central moment). Note that $\theta_1=0$ by definition, and $\theta_{2n}\theta_{2m}\leq\theta_{2n+2m}$ according to Cauchy-Schwarz inequality.

The value of $\theta_j$ depends on the characteristic variable $\epsilon \equiv \frac{v_n-\left\langle v_n\right\rangle}{\left\langle v_n\right\rangle}$. For a relatively narrow distribution for which the probability for $|\epsilon|>1$ is small, i.e $\int_{|\epsilon|>1} p(v_n)dv_n\ll1$~\footnote{In most cases, a weaker condition can be used: $\sigma_n\ll \left\langle v_n\right\rangle$, where $\sigma_n$ is the root-mean-square width.} and value of $j$ is not too large, we expect that $|\theta_j|\ll 1$ and $|\theta_j|$ decreases parametrically with increasing $j$. Plugging eq.~\ref{eq:5a} into the eq.~\ref{eq:3d}, and keeping terms to fourth order one obtains:
\begin{eqnarray}\nonumber
\mathcal{C}_n\{2\}/\left\langle v_n\right\rangle^2 &=& 1+\theta_2\\\nonumber
\mathcal{C}_n\{4\}/\left\langle v_n\right\rangle^4 &=& 1-2\theta_2-4\theta_3-\theta_4+2\theta_2^2\\\nonumber
\mathcal{C}_n\{6\}/\left\langle v_n\right\rangle^6 &\approx& 1-3\theta_2-4\theta_3+\frac{3}{2}\theta_4-\frac{9}{2}\theta_2^2\\\label{eq:5b}
\mathcal{C}_n\{8\}/\left\langle v_n\right\rangle^8 &\approx& 1-4\theta_2-\frac{56}{11}\theta_3+\frac{62}{33}\theta_4-\frac{40}{11}\theta_2^2
\end{eqnarray}
This leads to the following approximation for $v_2\{2k\}$:
\begin{eqnarray}\nonumber
v_n\{2\}/\left\langle v_n\right\rangle &\approx& 1+\frac{1}{2}\theta_2-\frac{3}{8}\theta_2^2\\\nonumber
v_n\{4\}/\left\langle v_n\right\rangle &\approx& 1-\frac{1}{2}\theta_2-\theta_3-\frac{1}{4}\theta_4+\frac{1}{8}\theta_2^2\\\nonumber
v_n\{6\}/\left\langle v_n\right\rangle &\approx& 1-\frac{1}{2}\theta_2-\frac{2}{3}\theta_3+\frac{1}{4}\theta_4-\frac{11}{8}\theta_2^2\\\label{eq:5c}
v_n\{8\}/\left\langle v_n\right\rangle &\approx& 1-\frac{1}{2}\theta_2-\frac{7}{11}\theta_3+\frac{31}{132}\theta_4-\frac{117}{88}\theta_2^2
\end{eqnarray}
Hence flow harmonics for all higher-cumulant $v_n\{2k\}$ are approximately the same~\cite{Voloshin:2007pc,Bilandzic:thesis}: 
\begin{eqnarray}
\label{eq:6a}
v_n\{2k\}\approx \left\langle v_n\right\rangle(1-\frac{1}{2}\theta_2), k>1.
\end{eqnarray}
The relative flow fluctuation can be obtained by combining second- and fourth-order cumulants, giving a well known result~\cite{Voloshin:2008dg}: 
\begin{eqnarray}
\label{eq:6b}
\frac{\sigma_n}{\left\langle v_n\right\rangle} \equiv \sqrt{\theta_2} \approx \sqrt{\frac{v_n^2\{2\}-v_n^2\{4\}}{v_n^2\{2\}+v_n^2\{4\}}} \equiv F_2
\end{eqnarray}
It can be shown that $F_2$ differs from $\sigma_n/\left\langle v_n\right\rangle$ by a factor $1+\frac{\theta_3}{2\theta_2}+\frac{\theta_4-\theta_2^2}{8\theta_2}+O(\theta_3)$. Lastly, eq.~\ref{eq:5c} also leads to two useful approximations that are valid for narrow $p(v_n)$ distributions.
\begin{eqnarray}
\nonumber
v_n\{6\}-v_n\{4\} &\approx& \left\langle v_n\right\rangle\left(\frac{1}{3}\theta_3 +\frac{\theta_4-3\theta_2^2}{2}\right)\\\label{eq:7a}
v_n\{8\}-v_n\{6\} &\approx& \left\langle v_n\right\rangle\left(\frac{1}{33}\theta_3 -\frac{\theta_4-3\theta_2^2}{66}\right)
\end{eqnarray}
The difference between $v_n\{6\}$ and $v_n\{8\}$ is about one order of magnitude smaller than the difference between $v_n\{4\}$ and $v_n\{6\}$. 

\section{Behavior of cumulants for broad distributions}
When the flow distribution is very broad, i.e. the probability of the events with $|v_n-\left\langle v_n\right\rangle|>\left\langle v_n\right\rangle$ is large (or $\sigma_n\gtrsim\left\langle v_n\right\rangle$), the higher-order central moments can no longer be treated as perturbation of the cumulants. All $\theta_j$ terms are in principle of the similar magnitude, and depending on the shape of the $p(v_n)$ the values of $c_n\{2k\}$ may differ significantly. We use the word ``may'' because for the well-known Gaussian distribution (eq.~\ref{eq:4}), the higher-order $c_n\{2k\}$ are always the same independent of the broadness of the distribution~\footnote{It is a broad distribution when $\delta_{n}\gg v_n^{\mathrm{0}}$, and a narrow distribution when $\delta_{n}\ll v_n^{\mathrm{0}}$.}. On the other hand, it is very easy to construct simple distributions for which $c_n\{2k\}$ differ from each other. This is an important point, because the $p(v_n)$ distribution of an ensemble can differ significantly from Gaussian, either for collisions obtained via event-shape engineering technique or for collisions of deformed nucleus. 

\begin{figure}[!h]
\center
\includegraphics[width=0.7\linewidth]{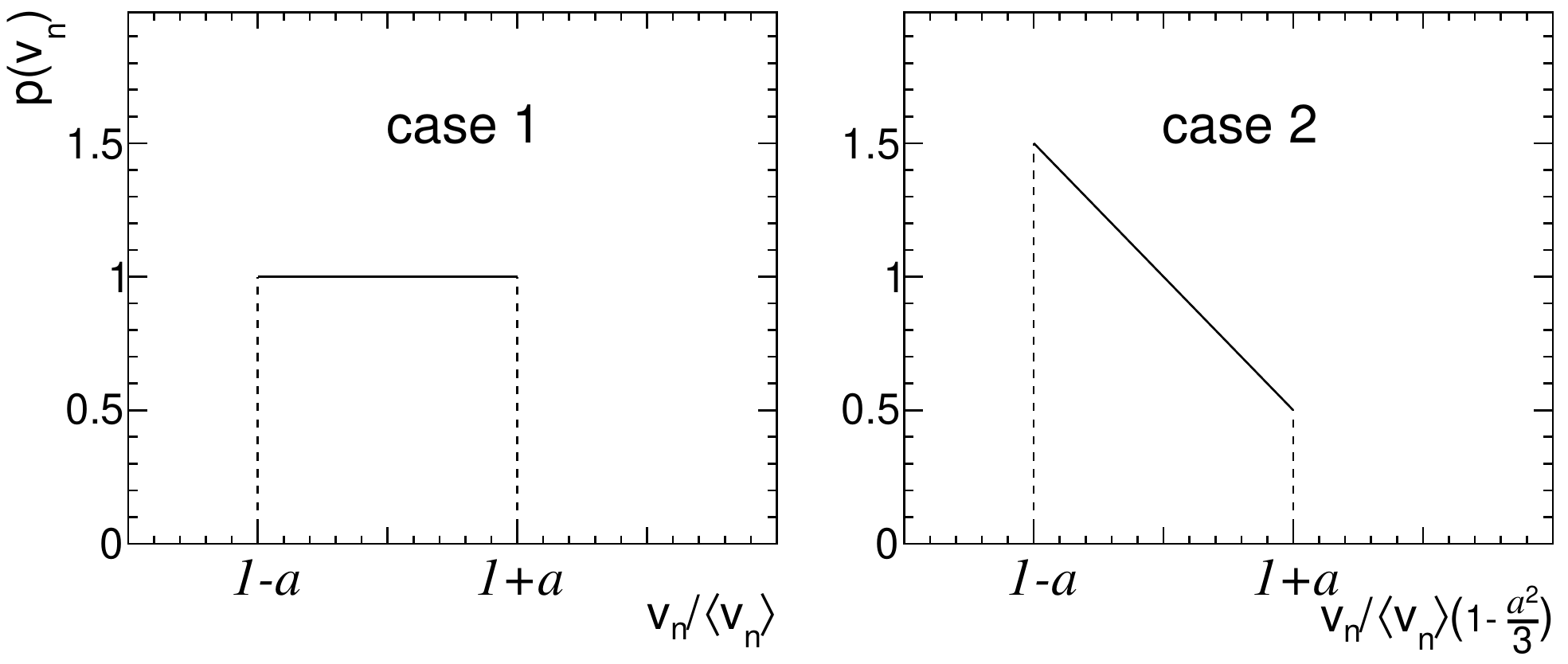}
\caption{\label{fig:1} The two flow probability distributions represented by eqs~\ref{eq:case1} and~\ref{eq:case2}. Parameter $a$ controls the width of these distributions. }
\end{figure}

In this section, two simple examples are discussed to illustrate the possible behavior of the cumulants when the associated distribution broadens. These distributions are shown in Fig.~\ref{fig:1} with the following expression:
\begin{eqnarray}
\label{eq:case1}
p(x;a) &=& \left\{\begin{array}{ll} 1 & |x-1|\leq a\\0 & |x-1|>a\\   \end{array}\right., \;\;\;\;\;\; x\equiv \frac{v_n}{\left\langle v_n\right\rangle}\\\nonumber\\\label{eq:case2}
p(x;a) &=& \left\{\begin{array}{ll} 2-x & |x-1|\leq a\\0 & |x-1|>a\\   \end{array}\right., x\equiv \frac{v_n}{\left\langle v_n\right\rangle}\left(1-a^2/3\right)
\end{eqnarray}
The width of the distributions increase with $a$, but their $\left\langle v_n\right\rangle$ values either remain the same (eq.~\ref{eq:case1}) or decrease (eq.~\ref{eq:case2}). Hence these examples are used to study one of the possible changing behavior of cumulants from $\sigma_n\ll \left\langle v_n\right\rangle$ to $\sigma_n>\left\langle v_n\right\rangle$.

\begin{figure}
\center
\includegraphics[width=0.45\linewidth]{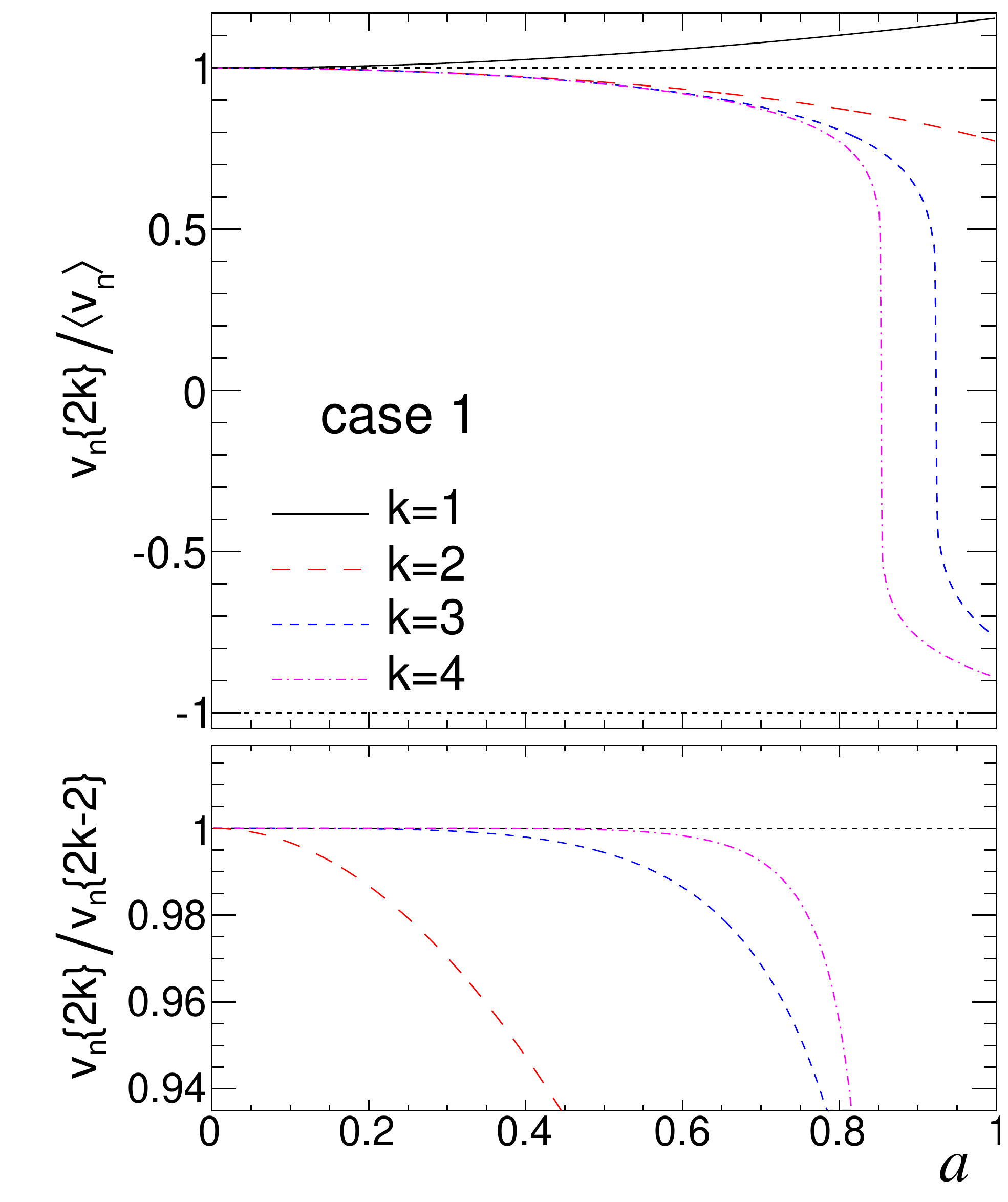}\includegraphics[width=0.45\linewidth]{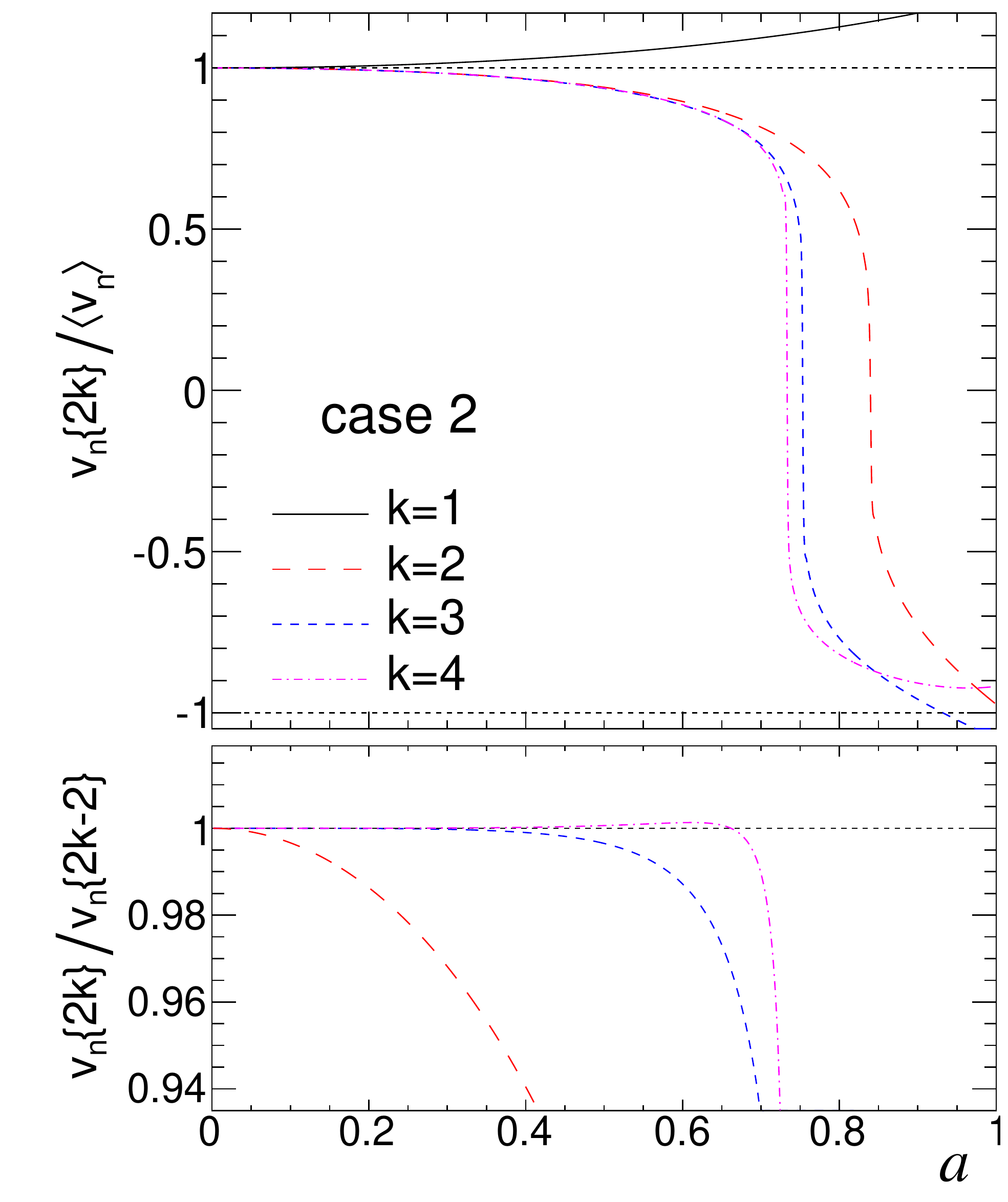}
\caption{\label{fig:2} The flow harmonics from cumulants $v_n\{2k\}$ (top panel) and (bottom panel) ratios between neighboring cumulant estimates $v_n\{2k\}/v_n\{2k-2\}$ as a function of $a$ for two probability distributions represented by Eq.~\ref{eq:case1} (left panel) and Eq.~\ref{eq:case2} (right panel).}
\end{figure}

The cumulants as a function of $a$ for these distributions are shown in Fig.~\ref{fig:2}. In both cases, $v_n\{6\}$ and $v_n\{8\}$ change sign for large values of $a$. In the second case, the value of $v_n\{4\}$ also become negative for $a\gtrsim0.84$. The transition happens rather abruptly due to the large exponential powers that relates $c_n\{2k\}$ and $v_n\{2k\}$. In the region where $v_n\{2k\}$ are positive (physical), the flow harmonics from higher-order cumulants are very close to each other since they are all dominated by $\left\langle v_n\right\rangle(1-\frac{1}{2}\theta_2)$, and hence are insensitive to the detailed shape of the underlying distributions. The differences are quantified by ratios between neighboring cumulants $v_{n}\{2k\}/v_{n}\{2k-2\}$. For both distributions, differences on the order of a few percents are observed between $v_{n}\{4\}$ and $v_{n}\{6\}$ for moderately large $a$ ($a>0.5$). The difference between $v_{n}\{6\}$ and $v_{n}\{8\}$ are always very small except close to the transition region ($a\gtrsim0.7$). 

The behavior of higher-order cumulants are often studied from the mathematical properties of the generation functions for flow cumulants~\cite{Bhalerao:2003xf}:
\begin{eqnarray}
\label{eq:c1}
-\ln\int J_0(2v_nz)p(v_n)dv_n = \sum_{k=1}^{\infty} \frac{z^{2k}a_k}{k!^2}\mathcal{C}_n\{2k\},\;\; v_n\{2k\}\equiv \mbox{sgn}(\mathcal{C}_n\{2k\})\sqrt[2k]{|\mathcal{C}_n\{2k\}|}
\end{eqnarray}
where the coefficients $a_k=1,1,4,33,456$... are defined by 
\begin{eqnarray}
\label{eq:c2}
-\ln J_0(2z) = \sum_{k=1}^{\infty} \frac{z^{2k}a_k}{k!^2}, 
\end{eqnarray}
One necessary condition for the convergence of $v_n\{2k\}$ for $k\rightarrow\infty$ is that ${f(z) = \int J_0(2v_nz)p(v_n)dv_n}$ must have zeros. However, this condition is not sufficient when $p(v_n)$ is non-Gaussian. This is because for a taylor series $\Sigma_{k=0}^{\infty} c_n z^k$, the radius of convergence $R$ is defined by its ``limit superior'': $\limsup\limits_{k\rightarrow\infty}\sqrt[k]{|c_n|} = R$, which is weaker requirement than the existence of $\lim\limits_{k\rightarrow\infty}\sqrt[k]{|c_n|}$.

\begin{figure}
\center
\includegraphics[width=0.7\linewidth]{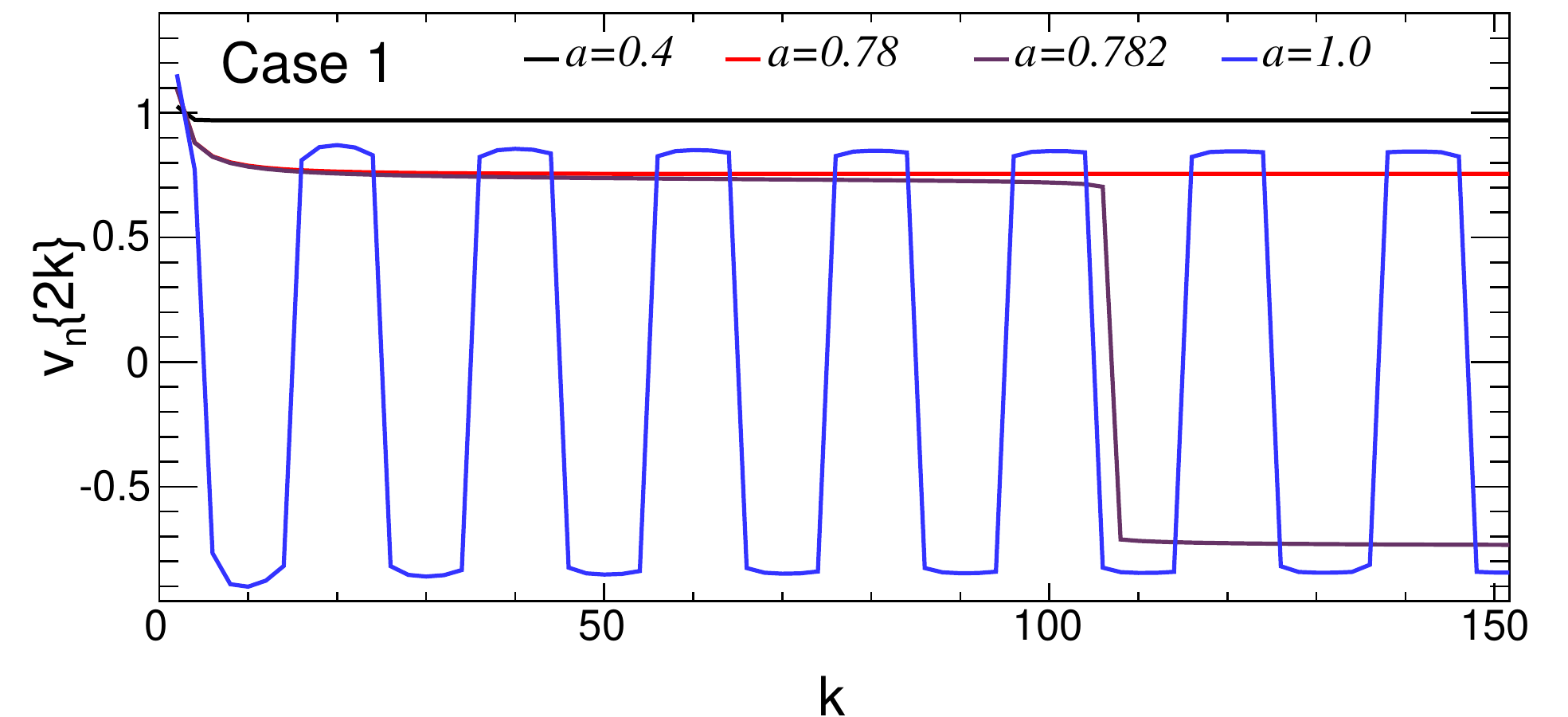}\vspace*{-0.92cm}
\includegraphics[width=0.7\linewidth]{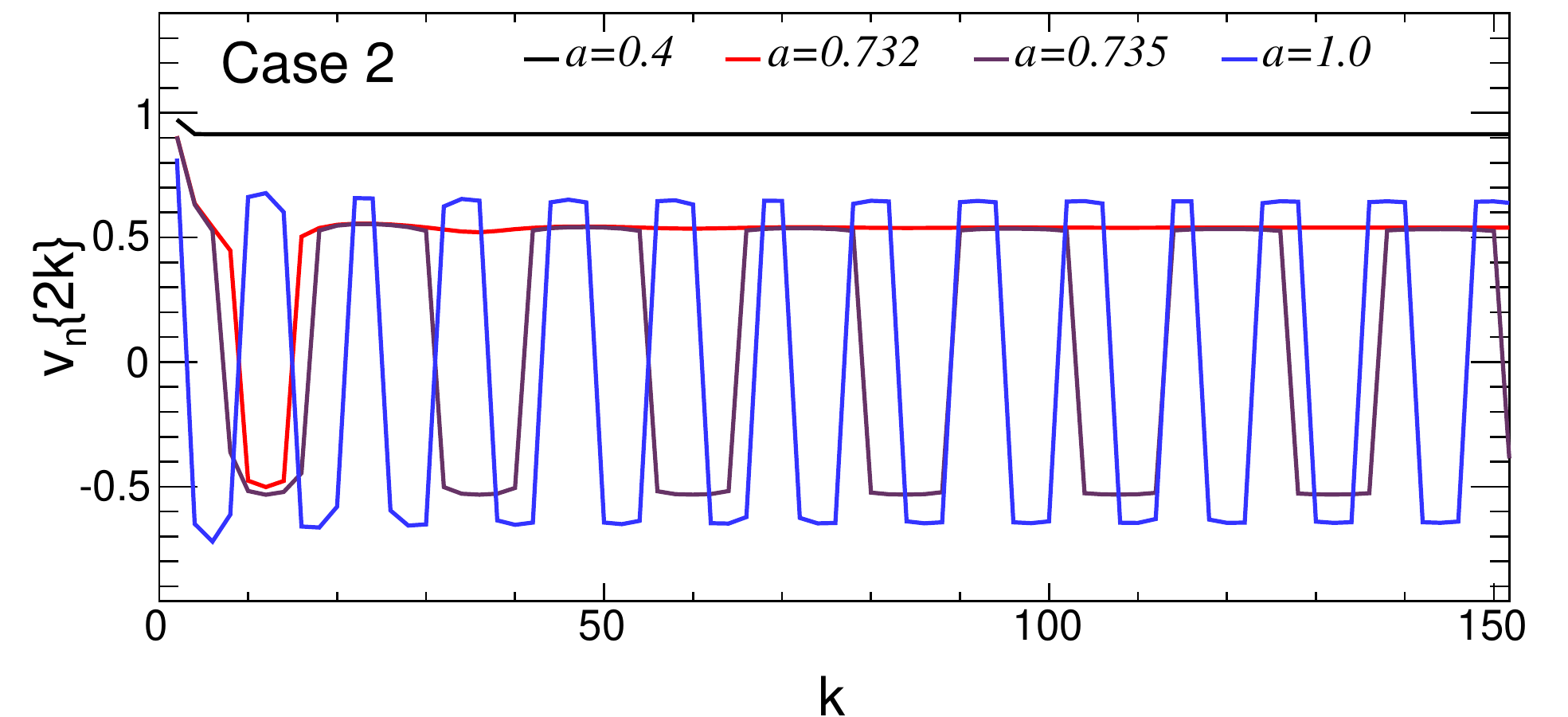}
\caption{\label{fig:2b} (Color online) The flow harmonics from cumulants $v_n\{2k\}$ as a function of $k$ for four values of $a$ for the two probability distributions represented by Eq.~\ref{eq:case1} (top panel) and Eq.~\ref{eq:case2} (bottom panel).}
\end{figure}

For the two distributions discussed above, we found that $v_n\{2k\}$ always converges for small $a$ (in general true for narrow distributions), however when $a$ is large,  Fig.~\ref{fig:2b} shows that the sign and magnitude of $v_n\{2k\}$ fluctuates. Clearly, their ``limit superior'' exist, but $v_n\{2k\}$ do not converge to a common value, and they do not even have the same sign. In this case, one need cumulants of all order to capture the shape of $p(v_n)$, which does not seem to be an efficient way to study flow fluctuations.

These examples illustrate the potential limitation of reconstructing the $p(v_n)$ with the flow harmonics estimated from higher-order cumulants. Two distributions with same $\left\langle v_n\right\rangle$ and same relative width $\theta_2$, but otherwise with very different shapes often have similar $v_n\{2k\}$ with a difference on the order of a few percent or less. Previous studies~\cite{Bzdak:2013rya,Yan:2013laa,Yan:2014afa} show this is also the case for other more physical distributions motivated by Glauber models in $p$+A or A+A collisions: these so call power or elliptic power functions have significant non-Gaussian tails but nevertheless very similar $v_n\{2k\}$. Distinguishing between these different distributions experimentally via flow harmonics obtained from higher-order cumulants thus require extreme precision and careful cancellation of the systematic uncertainties between $v_n\{2k\}$ of different order (see Fig.6 of Ref.~\cite{Yan:2014afa} and Fig.~9 of Ref.~\cite{Aad:2014vba}). Although the hierarchy between $v_n\{2k\}$ shows more sensitivity when $p(v_n)$ is very broad, this is also the region where all reduced central moments $\theta_j$ are equally important, and $v_n\{2k\}$ may change sign. These caveats need to be considered when inferring the shape of $p(v_n)$ from the measured $v_n\{2k\}$.

ATLAS Collaboration has obtained $v_2\{4\}$, $v_2\{6\}$ and $v_2\{8\}$ from the measured $p(v_2)$ in Pb+Pb collisions~\cite{Aad:2013xma}, they are found to be in excellent agreement with those measured directly via the cumulant methods~\cite{Aad:2014vba}. A significant non-Gaussianity of $p(v_2)$ is found to lead to a 1\%--2\% small difference between $v_2\{4\}$ and $v_2\{6\}$ in mid-central collisions, but no visible difference between $v_2\{6\}$ and $v_2\{8\}$. Our discussion above suggests that it may not be easy to reconstruct the non-Gaussianity of $p(v_2)$ from the directly measured $v_2\{4\}$, $v_2\{6\}$ and $v_2\{8\}$ within their respective experimental uncertainties.

Recently, a lot of studies have been devoted to the $v_n$ in $\pA$ collisions~\cite{CMS:2012qk,Abelev:2012ola,Aad:2012gla,Adare:2013piz,Aad:2013fja,Chatrchyan:2013nka,Aad:2014lta}, where the measurement of $v_2\{2k\}$ have been obtained for $k=1$ to 4. A sign change of $c_2\{4\}$ has been interpreted as transition from non-flow dominated region to flow dominated region (see Fig.~\ref{fig:0}(a)). But as discussed before, $c_2\{4\}<0$ in general does not necessarily indicate the onset of collectivity, unless we have a priori knowledge of its shape, e.g. it is close to Gaussian. Furthermore the fact that $v_2\{4\}\approx v_2\{6\}\approx v_2\{8\}$ in $p$+Pb collisions (see Fig.~\ref{fig:0}(b)) only implies a consistency with the dominance of collective behavior, and within current uncertainties it can not constrain the shape beyond its mean and width.

\section{``Event-by-event'' cumulant?}
After the measurement of $p(v_n)$ came out, there were suggestions to combine the advantage of cumulant method (good non-flow suppression) and EbyE $v_n$ method (more sensitive to flow distributions) by performing some kinds of event-by-event cumulant measurement. Strictly speaking, cumulants as well as moments should only be defined for an ensemble of events. But nevertheless, we could imagine calculating quantities analogous to cumulants for each event~\footnote{We choose the expression similar as Eq.~\ref{eq:3b} for direct analogy. In principle, statistical fluctuations may lead to non-zero odd moments within a single event~\cite{Bilandzic:2010jr}, which lead to additional higher-order corrections to Eq.~\ref{eq:ebe} but don't change the general conclusion.}:
\begin{eqnarray}\nonumber
\mathcal{C}'_n\{2\} &\equiv& \left\langle 2\right\rangle\\\nonumber
-\mathcal{C}'_n\{4\} &\equiv& \left\langle 4\right\rangle-2\left\langle2\right\rangle^2\\\nonumber
4\mathcal{C}'_n\{6\} &\equiv& \left\langle 6\right\rangle-9\left\langle4\right\rangle\left\langle2\right\rangle+12\left\langle 2\right\rangle^3\\
...&&\\\label{eq:8a}
v_n'\{2k\} &\equiv& \mbox{sgn}(\mathcal{C}'_n\{4\})\sqrt[2k]{|\mathcal{C}'_n\{2k\}|}
\end{eqnarray}
The formulae for EbyE multi-particle correlation are expressed in terms of ${\bf q}_n$ and $\omega\equiv 1/M$ following the direct cumulant framework of Ref.~\cite{Bilandzic:2010jr,Bilandzic:thesis}:
\begin{eqnarray}
\left\langle 2\right\rangle &=& \frac{q_n^2-\omega}{1-\omega}\\
\left\langle 4\right\rangle &=& \frac{q_n^4-4\omega q_n^2\left[(1-2\omega)+\frac{1}{2}q_{2n}\cos2n(\Psi_{2n}-\Psi_n)\right]+2\omega^2\left[1-3\omega+\frac{1}{2}q_{2n}^2\right]}{(1-\omega)(1-2\omega)(1-3\omega)}\\\nonumber
\left\langle 6\right\rangle &=& \left(q_n^6-9\omega q_n^4\left[(1-4\omega)+\frac{2}{3}q_{2n}\cos2n(\Psi_{2n}-\Psi_n)\right]\right.\\\nonumber
&&+18\omega^2q_n^2\left[(1-2\omega)(1-5\omega)+(1-4\omega)q_{2n}\cos2n(\Psi_{2n}-\Psi_n)+\frac{1}{2}q_{2n}^2+\frac{2}{9}q_{3n}q_{n}\cos 3n(\Psi_{3n}-\Psi_n)\right]\\\label{eq:ebe}
&&\left.-6\omega^3\left[(1-4\omega)(1-5\omega)+\frac{3}{2}(1-4\omega)q_{2n}^2+2q_nq_{2n}q_{3n}\cos n(3\Psi_{3n}-2\Psi_{2n}-\Psi_n)\right]+4\omega^4q_{3n}^2\right)\left.\vphantom{\frac{|}{|}}\middle/\Pi_{k=1}^{5} (1-k\omega)\right.
\end{eqnarray}
The formula for $\left\langle 8\right\rangle$ is skipped since it is quite lengthy. The terms in these formula are ordered in powers of $\omega$. The higher-order terms in the 2k-particle correlation account for contributions from combinations where some angles in Eq.~\ref{eq:3a} are identical (or duplicates). These terms are can be large when $\omega \sim q_n^2$, and hence they are important for flow moments. However the cumulant definition naturally suppresses these duplicates, for example one can show that:
\begin{eqnarray}\nonumber\\
\mathcal{C}'_n\{4\}&\approx& q_n^4+2\omega q_n^2\left(4\omega+q_{2n}\cos2n(\Psi_{2n}-\Psi_n)-q_n^2\right)-2\omega^3\\
\mathcal{C}'_n\{6\}&\approx& q_n^6+3\omega q_n^4\left(3\omega+q_{2n}\cos2n(\Psi_{2n}-\Psi_n)-q_n^2\right)
\end{eqnarray}
For sufficiently large multiplicity such that $q_n \gg\omega$ or $Mq_n\gg1$, event-by-event cumulants $\mathcal{C}'_n\{2k\}$ should be dominated by the leading term in each event: $\mathcal{C}'_n\{2k\}\approx q_n^{2k}$ and $v'_n\{2k\}\approx q_n$~\footnote{The influence of high-order terms are more important for moments. Only when $q_n^2 \gg\omega$ or $Mq_n^2\gg1$,  $\left\langle 2k\right\rangle \approx q_n^{2k}$.}. In this case, the non-flow contributions to $v'_n\{2k\}$ is nearly completely controlled by their contributions to $q_n$.

\begin{figure}
\center
\includegraphics[width=0.8\linewidth]{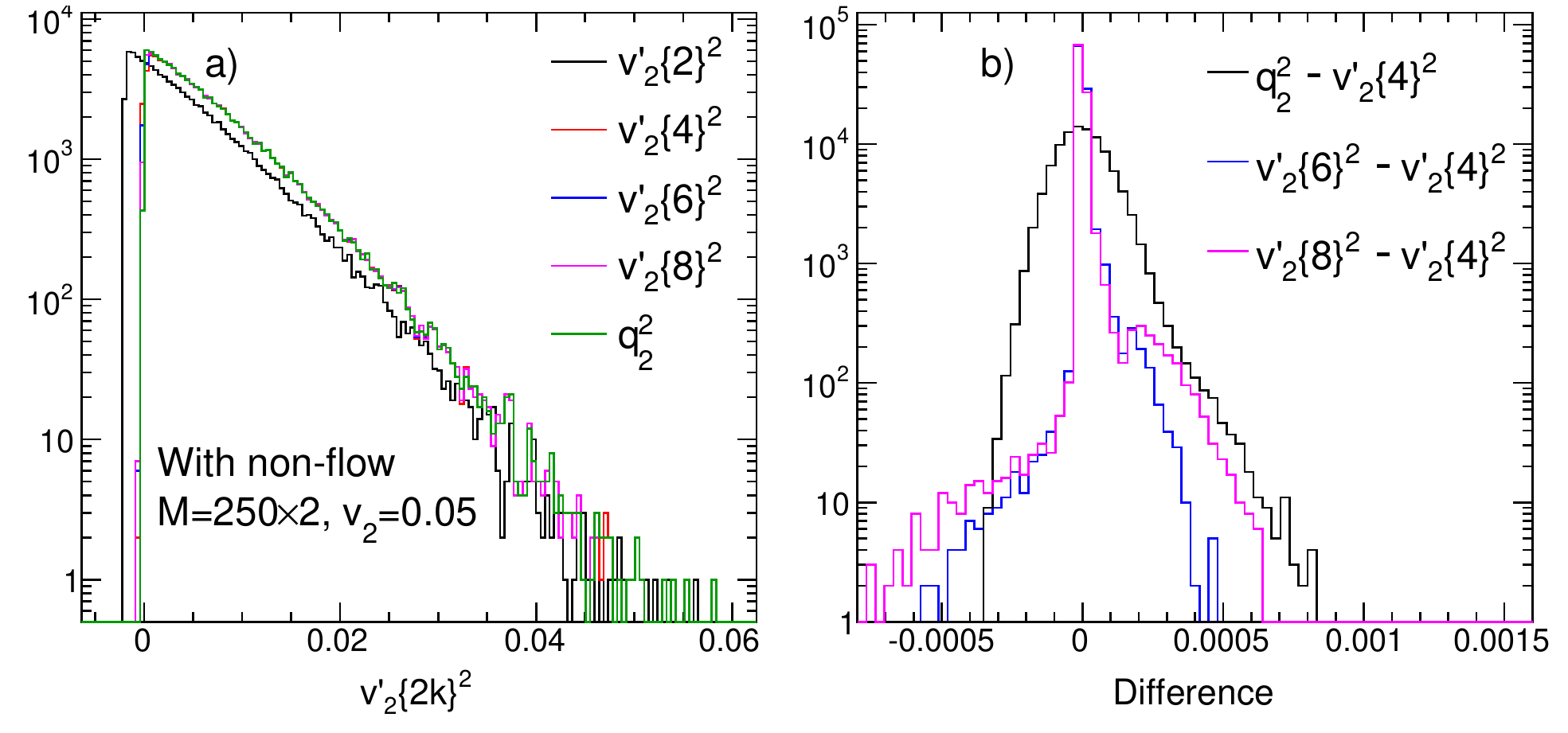}
\caption{\label{fig:5} (Color online) (a) The event-by-event distributions of $v_n'\{2k\}^2$ defined by eq.~\ref{eq:9} and $q_n^2$ from the toy simulation. (b) The event-by-event difference between these variables.}
\end{figure}
To verify this, a simple toy simulation is performed with $M$=500 and $v_2=0.05$. The 500 particles are generated from 250 resonances each producing one pair of particles in the same direction, so the non-flow effects from statistical fluctuation and resonance decays have similar magnitudes. The results are calculated as
\begin{eqnarray}
\label{eq:9}
v_2'\{2k\}^2 \equiv \mbox{sgn}(\mathcal{C}'_2\{2k\})\sqrt[k]{|\mathcal{C}'_2\{2k\}|}
\end{eqnarray}
and they are shown in Fig.~\ref{fig:5}(a) (in this definition, $v_2'\{2\}^2$ and $q_2^2$ are related by a simple linear transformation). The fluctuations of $v_2'\{2k\}^2$ reflect statistical fluctuation and non-flow from resonance decays and they are nearly identical as expected (also reflected by the EbyE difference between the higher-order EbyE cumulants shown in Fig.~\ref{fig:5}(b)). This result is consistent with the fact that non-flow suppression of the cumulants is achieved by averaging over events, not within each event.
\section{Summary and discussion}
The relationship between the cumulants and the event-by-event flow distribution $p(v_n)$ has been investigated via a central moment expansion approach. For a narrow distribution where the width is much smaller than the mean, i.e $\sigma_n\ll \left\langle v_n\right\rangle$, the flow harmonics from higher-order cumulants $v_n\{2k\}$ for $k>1$ are very close to each other: $v_n\{2k\}\approx \left\langle v_n\right\rangle\left(1-\sigma_n^2/ \left\langle v_n\right\rangle^2\right)$. Thus similarity of flow harmonics from higher-order cumulants for elliptic flow, $v_2\{4\}\approx v_2\{6\}\approx v_2\{8\}$, in A+A collisions and high-multiplicity $p$+A collisions does not provide strong constraints on the shape of $p(v_2)$ beyond its mean and width. As the distribution becomes broader $\sigma_n\gtrsim \left\langle v_n\right\rangle$,  depending on the shape of $p(v_n)$, the $v_n\{2k\}$ may start to differ from each other, and eventually become negative. This sign change arises from the choice of $p(v_n)$, without the need to invoke non-flow effects. Hence the sign change of $c_2\{4\}$ at small multiplicity, in principle does not have to indicate the onset of collectivity or non-flow, without some a priori assumption of the shape of $p(v_2)$ based on Glauber model. 

For these reasons and also because cumulants only probes the even moments of the $p(v_n)$, direct measurement of $p(v_n)$ via a data-driven unfolding method is a more preferred approach, which was shown to directly remove statistical fluctuation and largely suppress various short-range correlations~\cite{Jia:2013tja}. The shape-resolving power of this method is limited only by the width of the EbyE non-flow contribution not by $\sigma_n/\left\langle v_n\right\rangle$. Furthermore, cumulants can be calculated directly from the measured $p(v_n)$, typically with smaller systematic uncertainties than direct calculations~\cite{Aad:2014vba}. 

In fact, if all one cares about are $v_n\{2k\}$, they can be estimated directly from $p(q_n)$ without need of unfolding in the following way. Since the $p(v_n)$ is close to Bessel-Gaussian (Eq.~\ref{eq:4a}), additional smearing from statistical fluctuation and other non-flow effects mainly increase $\delta_n$ but have little impact on $v_n^0$. Assuming the width of non-flow distribution $p({\boldsymbol s}_{n})$ (including the statistical fluctuations) is $\delta_{\mathrm{nf}}$ in the $x$- or $y$-axis direction, then it is easy to see that the flow harmonics can be estimated from $p(q_n)$ as
\begin{eqnarray}
\label{eq:10}
&&v_n\{2\}^2 \approx q_n\{2\}^2-2\delta_{\mathrm{nf}}^2,\\
&&v_n\{2k\}\approx q_n\{2k\} \approx v_n^0, k>1
\end{eqnarray}
where $q_n\{2k\}$ are calculated directly from $p(q_n)$ using formula identical to Eqs.~\ref{eq:3d} and \ref{eq:3e}. Using HIJING events generated with realistic flow afterburner~\cite{Jia:2013tja}, we have verified that $q_n\{2k\}$ is essentially the same as the $v_n\{2k\}$ obtained using the direct cumulant method~\cite{Bilandzic:2010jr} for central and mid-central Pb+Pb collisions at RHIC and LHC energies~\cite{more}.

We also investigated the statistical nature of cumulant calculated on an event-by-event bases. It was shown that these EbyE cumulants are nearly completely determined by observed flow vector $q_n$, and hence they provide no benefit in suppressing statistical smearing and other non-flow effects on a EbyE bases.

We appreciate valuable comments and fruitful discussions with D.~Teaney and A.~Bilandzic. This research is supported by NSF under grant number PHY-1305037.
\bibliography{cumulantv6}{}
\bibliographystyle{apsrev4-1}
\end{document}